\documentstyle[editedbook,epsfig,psfig,epsf]{mq}

\def\cm2{cm$^2$ }
\def\se1{s$^{-1}$ }

\begin{opening}
\title{The 2002 outburst of the microquasar XTE J1550-564}
\author{A. P. Colombo$^{1,2}$,  T. Belloni$^{2}$, J. Homan$^{2}$, S. Campana$^{2}$, M. van der Klis$^{3}$}
\institute{$^1$ University of Insubria, Como, Italy \\
	   $^2$ INAF -- Osservatorio Astronomico di Brera, Merate, Italy\\
	   $^3$ Astronomical Institute, Univ. of Amsterdam, the Netherlands}
\end{opening}

\runningtitle{The 2002 outburst of the microquasar XTE J1550-564}
\runningauthor{A. P. Colombo, T. Belloni, J. Homan, S. Campana \& M. van der Klis}
\begin{document}
\vspace{-0.5cm}
\begin{abstract}
{\small We present results of spectral and timing analysis based on 11 RXTE PCA/HEXTE observations of the microquasar XTE J1550-564 during its last outburst in January 2002. 
The observed behaviour is comparable to most Black-Hole Candidates in the low/hard state This is unlike the 1998-99 outburst, when it showed a much more complex feature, probably because of the higher luminosity. For each of the 11 observations we extracted energy spectra and power density spectra, finding typical features of a low/hard state.\\ }
\end{abstract}

\section{Introduction}
The microquasar XTE J1550-564 was discovered with the All Sky Monitor on board RossiXTE \cite{levine96} in September 1998 \cite{smith98} and to date it showed four outbursts. The compared analysis of all these outbursts allows to relate different behaviours of the source with different values of flux reached during the outburst. The outburst here analyzed is quite different from the first two and similar to the 2001 one. Radio observations were reported by \cite{corbel02}.
A more complete analysis is described in \cite{belloni2002}.

\section{Energy spectra}
We analyzed spectra in a range from 3 to 150 keV from PCA \cite{jahoda96} and HEXTE \cite{rothschild98} instruments corresponding to channels 4-52 for the PCA and 10-50 for HEXTE. 
We used a model consisting of a power law with a high-energy cutoff, a smeared Fe edge and a Fe emission line. The hydrogen column density was fixed at 8.5 $\times$ 10$^{21}$ cm$^{-2}$ and a systematic error of 0.75\% was quadratically added to the data. For the first seven observations the line was not necessary while for the remaining the edge was not required. The absence of a soft thermal component may be due to the lack of data at energies lower than 3 keV. The unabsorbed 3-150 keV flux decreased from 1.2 $\times$ 10$^{-8}$ erg cm$^{-2}$ s$^{-1}$ to 7.1 $\times$ 10$^{-10}$ erg cm$^{-2}$ s$^{-1}$, corresponding to a luminosity at 6 kpc of 5.0-0.2 $\times$ 10$^{37}$ erg s$^{-1}$. 

The values found for the power law photon index are tipical of a low/hard state. The cutoff energy decreases substantially during the observations (from $\sim$340 keV to $\sim$64 keV) while the photon index does not increase much (from $\sim$1.4 to $\sim$1.5).

\begin{figure}[htb]
\centering
\psfig{file=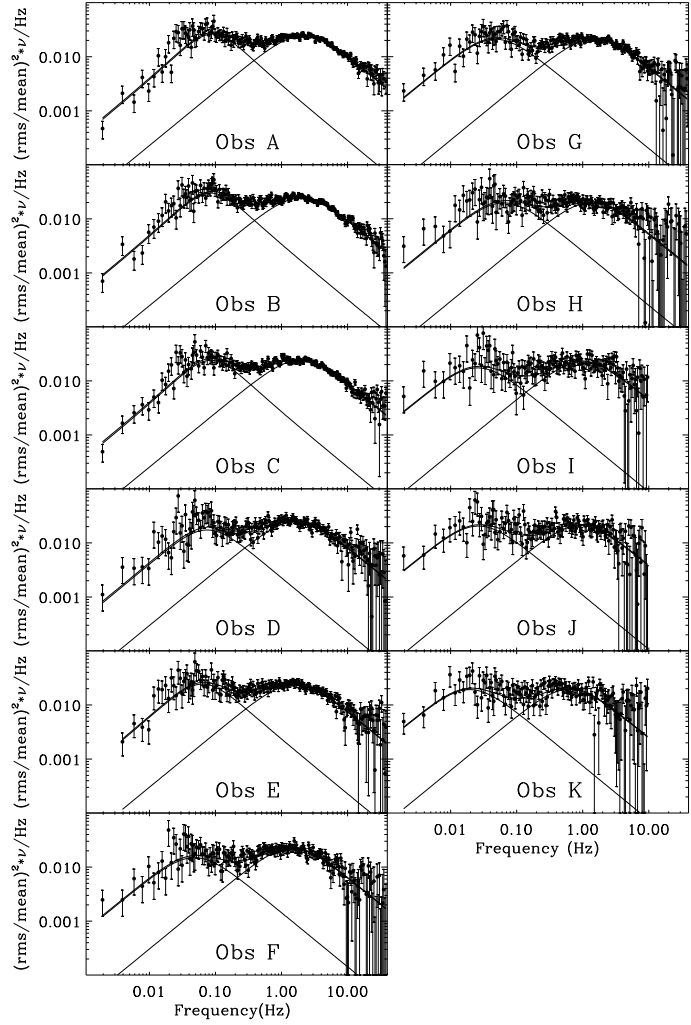,width=10cm}
\caption{Power Density Spectra for all the observations presented here in $\nu$P$_{\nu}$ form. The best fit models are also shown.}
\label{powspe}
\end{figure}

\section{Power density spectra}
The variability observed in the PCA light curve is tipical of the low/hard state. For each observation we created a Power Density Spectrum (PDS) in the following way: we produced individual Leahy-normalized \cite{lehay83} PDS from 512-second stretches with time resolution of 512 s$^{-1}$ from PCA data ranging from channel 0 to 35 (roughly corresponding to energies from 2 to 15 keV). These PDS were averaged and the contribution due to Poisson statistic was subtracted \cite{zhang95}. The averaged PDS were normalized to squared fractional rms \cite{bellonihasinger90a}. 

Each PDS was fitted with a model consisting of a sum of Lorentzian components \cite{bpk02}: we always needed only two broad Lorentzian to reach good values of the chi-squared. In Figure ~\ref{powspe} all the PDS extracted are shown, together with their best fit Lorentzian components.

\section{Discussion}
From the analysis of energy spectra we can see how the presence of a high energy cutoff in the power law component is consistent with the presence of a thermal distribution of electrons in the inner region. In such a model the decrease of the cutoff energy implies a decrease of electron temperature: since the photon index does not change correspondingly the optical depth of the Comptonizing cloud must have increased at the same time.

Interpreting the two Lorentzian components present in the PDS as L$_b$ and L$_{\ell}$ (see \cite{bpk02}), we can verify how the values found are consistent with existing correlations found in other sources, like the one concerning $\nu_b$ vs. $\nu_{\ell}$ \cite{straaten02} and $\nu_b$ vs. rms$^2$ \cite{bellonihasinger90b}. The timing analysis results confirm that XTE J1550-564 is in a low/hard state. 

The general behaviour showed by XTE J1550-564 during this outburst is compatible with the one observed in other BHC in their low/hard state, indicating that at lower fluxes (and therefore possibly lower accretion rates) the source behaves like ``normal'' BHC.




\end{document}